\documentclass{mn2e}
\usepackage{graphicx}
\usepackage{amssymb}

\title[TRGB Distance Determinations] {Determining the Location of the Tip of the Red Giant Branch in Old Stellar Populations: M33, Andromeda I \& II.}
\author [McConnachie et al.] 
{A. W. McConnachie${^1}$, M. J. Irwin${^1}$, A. M. N. Ferguson${^2}$, R. A. Ibata${^3}$,\newauthor
G. F. Lewis${^4}$, N. Tanvir${^5}$\\  
${^1}$ Institute of Astronomy, Madingley Road, Cambridge, CB3 0HA, UK\\
${^2}$ Max-Planck-Institut f\"{u}r Astrophysik, Karl-Schwarzschild-Str. 1, Postfach 1317, D-85741 Garching, Germany\\
${^3}$ Observatoire de Strasbourg, 11, rue de l'Universite, F-67000, Strasbourg, France\\
${^4}$ Institute of Astronomy, School of Physics, A29, University of Sydney, NSW 2006, Australia\\
${^5}$ Physics, Astronomy and Mathematics Dept., Univ. of Hertfordshire, Hatfield, AL10 9AB, UK\\}

\begin{document}

\maketitle

\begin{abstract}

The absolute bolometric luminosity of the point of core Helium ignition in old, metal poor, red-giant stars is of roughly constant magnitude, varying only very slightly with mass or metallicity. It can thus be used as a standard candle. Here, we review the main difficulties in measuring this location in any real dataset and develop an empirical approach to optimise it for tip of the red giant branch (TRGB) analysis. We go on to present a new algorithm for the identification of the TRGB in nearby metal poor stellar systems. Our method uses a least-squares fit of a data-adaptive slope to the luminosity function in $1^m$ windows. This finds the region of the luminosity function that shows the most significant decline in star counts as we go to brighter magnitudes; the base of this decline is attributed as the location of the tip. This technique then allows for the determination of realistic uncertainties which reflect the quality of the luminosity function used, but which are typically $\sim 0^m.02$ rms + $\sim 0^m.03$ systematic, a significant improvement upon previous methods that have used the tip as a standard candle. Finally, we apply our technique to the Local Group spiral galaxy M33 and the dwarf galaxies And I \& II, and derive distance modulii of $24^m.50 \pm 0^m.06$ ($794 \pm 23$ kpc), $24^m.33 \pm 0^m.07$ ($735 \pm 23$ kpc) and $24^m.05 \pm 0^m.06$ ($645 \pm 19$ kpc) respectively. The result for M33 is in excellent agreement with the Cepheid distances to this galaxy, and makes the possibility of a significant amount of reddening in this object unlikely.

\end{abstract}


\section{Introduction}

The accurate determination of distances in the Universe is a fundamental and difficult problem in astronomy. Parallax is the ideal method of distance determination as it is completely independent of the physical nature of the body of interest. However, this geometrical method can only be currently applied successfully in the nearby Galaxy due to limitations in the accuracy of the astrometry required. For systems external to this we are forced to search for some standard candle - an object whose intrinsic brightness we think we know. Measurement of the apparent brightness can then, after correction for extinction effects, lead to an estimate of its actual distance. The most successful and reliable standard candle known in astrophysics is the Cepheid variable, whose well studied Period-Luminosity relation sets the foundation for the entire cosmological distance ladder (see, for example, Tanvir 1999).

One of the limitations of the Cepheid method is that it is confined to relatively massive Population I systems. In order to calculate distances to older, more metal poor galaxies some other standard candle is required. The Tip of the Red Giant Branch (TRGB) offers an ideal alternative for these systems. Physically, this stage in stellar evolution represents the point of the core helium flash. Here, the temperature of the degenerate, quasi-isothermal core is only dependent upon the properties of the thin H-burning shell around it and this varies only very slightly with chemical abundance and surrounding mass. The TRGB is thus of roughly constant intrinsic brightness. Iben \& Renzini (1983) realised that for low mass, metal poor stars (${\rm [Fe/H]} < -0.7$ dex) the TRGB varies by only $\sim 0^m.1$ for any star older than 2 Gyrs. The absolute I band magnitude is also found to be constant for these stars as absorption effects in the stellar atmosphere are minimal, but for stars more metal rich than this molecular absorption causes the TRGB to appear dimmer in this band. A more recent study of the theoretical predictions concerning this stage in stellar evolution has been conducted by Salaris \& Cassisi (1997) and a comparison with other distance indicators, such as Cepheid and RR Lyrae variables, was made. They found agreement between these methods at the level of $\sim 0^m.1$.

Da Costa \& Armandroff (1990) compared the brightest RGB stars in various globular clusters to the theoretical predictions for core helium ignition and found very good agreement between the two, with only a very weak dependency on metallicity. They were dealing with relatively sparse colour magnitude diagrams however. A more recent study by Bellazzini, Ferraro \& Pancino (2001) of the globular cluster $\omega$ Centauri came to similar conclusions. They obtained a value of the absolute I band magnitude of the TRGB of $-4^m.04 \pm 0^m.12$. We shall later use this value in order to calculate the distance modulus to several galaxies (see Section 3.2 for a discussion of this result).

Several authors have used various techniques to derive TRGB distances. For example, Harris et al. (1998) used a model fitting technique to the luminosity function, although the most influential method to determine the location of the TRGB is due to Lee et al. (1993). Prior to this publication, estimates of the location of the TRGB had mostly been made by eye. Such a qualitative approach is of course not ideal as it is not reproducible. However, Lee et al. (1993) put the method on a quantitative basis by employing an edge detection algorithm. Essentially, an edge detection algorithm in this context compares neighbouring star counts and looks for the largest absolute change between neighbouring bins - in essence, it models the TRGB edge as a step. Specifically, Lee et al. used a zero-sum Sobel kernel [-2, 0, 2] and convolved it with the binned luminosity function. The range over which this kernel is applied is defined by the sampling of the luminosity function. In practise, the output of this displays a maximum in the bin coincident with the largest absolute change in the counts. The midpoint of this bin was then assumed to be the magnitude of the TRGB. Typical uncertainties in this quantity were quoted as $0^m.1 - 0^m.2$. 

Three years later, Sakai et al. (1996) adapted this method to make it independent of binning. They constructed a smoothed luminosity function $\wp(m)$ such that

\begin{equation}
\label{smooth}
\wp\left(m\right)=\sum_{i=1}^{N_\star} \frac{1}{\sqrt{2\pi}\sigma_i}\exp\left(-\frac{\left(m_i-m\right)^2}{2\sigma_i}\right),
\end{equation}

\noindent where $m_i$ and $\sigma_i$ are the magnitude and photometric error of the $i^{th}$ star from a sample of $N_\star$. In the limit of large $N_\star$, the result can be thought of as a luminosity probability distribution. Hereafter, we adopt the abbreviation LPD for this function, to distinguish it from the binned luminosity function. Sakai et al. then convolved this with a smoothed Sobel Kernel, this time in the form of [-2, -1, 0, 1, 2] - this kernel is less sensitive to spikes due to Poisson noise than the previously used one, as it allows for some smoothing of the luminosity function over a small range. This allowed them to find the position of the TRGB, as they had defined it, to within an uncertainty of $\sim 0^m.1$. More recently, Mendez et al. (2002) have refined this method further, and they have also used it in conjunction with a maximum likelihood method. This involved modelling the RGB in the luminosity function as a truncated power law and attributing the TRGB as the location at which the power law is truncated. We believe that such a technique is a significant improvement over the Sobel edge-detection algorithms and it offers many advantages - for example it is much more stable against noise effects. However, a step model (an abrupt cut-off in number counts) is still implicitly assumed as the intrinsic shape of the TRGB in their analysis.

It is not clear that an edge detection algorithm, or any method that assumes a step in the luminosity function, is necessarily the best approach to this problem, as there is no a priori reason why the luminosity function at the location of the tip should be a simple step function. It may equally well be a slope, or a parabola, or something else - the exact shape of the luminosity function at the TRGB edge is currently unknown. We note that Zocacali \& Piotto (2000) show that the bright end of RGB luminosity function is well modelled by a simple power law, but there is no model that predicts the shape of the TRGB edge, although this doubtlessly depends upon the number of stars that contribute to the luminosity function. Generally, we might expect the number of stars to decrease {\it gradually} over a short magnitude range as we approach the tip from along the RGB. The luminosity function would then exhibit more of a slope at the TRGB, an abrupt change in the bright end of this slope would then be, by definition, the TRGB. The Sobel edge detection algorithm does not always produce a peak in such a situation (see, for example, Figure 1 of Madore \& Freedman 1995). In practice, the resulting signal is not easy to detect over and above all the other random variations that the output of this procedure creates when dealing with all but the most idealised noise-free luminosity functions. 

In Section 2 of this paper, we review the main difficulties with using the TRGB as a distance indicator. We then describe and test a new algorithm that we have created in order to measure the TRGB location. This algorithm is based upon a least-squares fitting procedure and makes less assumptions about the shape of the luminosity function at the TRGB, working equally well for situations where the TRGB is marked by a sudden or gradual rise in star-counts. This data adaptive slope technique gives uncertainties in the distance modulus obtained typically of order $\sim 0^m.05$, but most importantly reflects the quality of the luminosity function used. We also demonstrate an additional simple empirical heuristic scheme that allows a first pass estimate of this location to be made. In Section 3 we go on to apply our method to calculate the distances to the spiral galaxy M33 and the dwarf galaxies And I \& II.

\section{Locating the TRGB}

\subsection{Data Considerations}

Finding the apparent magnitude of the TRGB in a satisfactory, quantitative manner has proved a challenge with several different techniques having been used in the literature. Madore \& Freedman (1995) have already written an excellent review and used computer simulations to analyse the effects of these difficulties on the Sobel edge detection algorithm. They concluded:

\begin{enumerate}
\item low signal to noise (S/N) can hide the location of the tip in a luminosity function or lead to the false identification of this point, with a jump due to noise possibly being attributed to the luminosity of core He ignition;
\item bright Asymptotic Giant Branch (AGB) stars can contaminate the luminosity function in the region of the tip, and may hide its true location;
\item foreground stars masquerade as members of the stellar system and will contaminate the luminosity function. 
\end{enumerate}

Poisson noise is usually the dominant uncertainty in any luminosity function, and becomes a serious issue if we intend to try and measure a specific location defined by some discontinuity in star-counts. Although the algorithm we describe later is less sensitive to Poisson noise than an edge detection algorithm, or such like, it is still not immune. The only way to overcome Poisson errors successfully is to get photometry for more stars. As well as decreasing the resulting noise in a luminosity function, large star-counts also ensure that the luminosity function we construct from our sample is as close to the underlying luminosity function of the population as possible. Therefore, the most luminous RGB stars are more likely to be at the point of core helium ignition - although we note that Crocker \& Rood (1984) conducted Monte-Carlo simulations which suggest that there is a large probability of observing a star close to the tip location even for a sample of relatively few RGB stars. Additionally, the effect of Malmquist bias on a well sampled luminosity function is negligible.

Many previous TRGB distance studies have used Hubble Space Telescope (HST) WFPC2 data (eg. Kim et al. 2002, Maiz-Appellaniz et al. 2002). However, despite WFPC2's superb resolution, its relatively small field of view (an L-shaped 2.5' x 2.5') may sometimes not allow for as large a number of bright RGB stars to be observed as is required for a relatively noise-free luminosity function suitable for TRGB analysis. Instead, wherever possible, to maximise the number of stars observed and to increase the reliability of the TRGB measurement, we advocate the use of ground based wide field cameras, for example the Wide Field Camera on the Isaac Newton Telescope ($0.29^\circ$ FOV) or MEGACAM on the Canada-France-Hawaii Telescope ($1^\circ$ FOV). The accuracy of the technique that we introduce later is significantly improved if used on a well-defined luminosity function, although realistic error estimations are obtained regardless of the number of stars contributing to the measurement. In these situations however, we may need to concern ourselves with whether the luminosity function we create is representative of the underlying population.

Having maximised the number of stars with reliable photometry, we then want to increase the S/N  for the resultant luminosity function. By signal we refer to RGB stars in the system of interest, and by noise we refer to everything else - AGB stars and foreground stars, in particular. The most obvious way to increase the fraction of RGB stars on the luminosity function is to take the luminosity function along the red giant branch only. This is most easily illustrated by looking at Figure 8, which shows the colour magnitude diagram for And I \& II. The luminosity function is only plotted for the stars bounded by the dotted lines. Their positions are determined by taking star counts in strips parallel to the RGB and placing the boundaries on either side of the maximum. The exact position or separation of the boundaries makes no difference to the derived value for the tip, as should be expected. All the other stars outside of this area are ignored, as they contribute nothing to the RGB and would only act to hide the tip. Some bright AGB stars may still be present but their effect has been lessened. Further, as we demonstrate in Section 2.2.1, our technique for finding the RGB appears robust even with a significant AGB population.

Some foreground stars will have survived the cut that we have applied to our data and will be contaminating our RGB luminosity function. We cannot hope to identify foreground stars individually but with wide field data we are able to apply a statistical correction for them by using a luminosity function of a neighbouring region consisting only of foreground stars. Then, assuming all the brighter stars in our fields of interest are foreground objects, we can use these to scale the reference luminosity function to our data if necessary.

Finally, we propose to use, in the same way as Sakai et al. (1996), the LPD given by Equation 1 instead of the normal binned histogram. This makes the process independent of any binning considerations and yields a continuous function for the analysis, with the drawback that the error analysis becomes more complicated than for a normal luminosity function. Taken as a whole, the above methodology creates clean LPDs optimised to measure the TRGB. Further specific discussion of our data is deferred until Section 3 after we have describing the algorithm we are using to analyse LPDs for TRGB locations. 

\subsection{Least-Squares Fitting of an Adaptive Slope}

The technique that we use to locate the TRGB involves finding the region on a LPD that shows the most significant decline in star counts. The base of this decline (slope) will be attributable to the TRGB. The main complication arises due to our lack of knowledge of the intrinsic shape of the LPD around the TRGB location, and what gradient best resembles the slope in this region. Fitting an optimum model to the LPD therefore implies a generalised least-squares fitting procedure. Note that the model we use is not defined over the full range of the LPD, as it is not our intention to model the LPD but to locate a specific point of the LPD. By making this our goal we minimise the assumptions that we are forced to make concerning the intrinsic shape of the LPD: we instead repeatedly look at the LPD through a window, and find the window in which the LPD is best fit as a simple slope function, that is, the region that when coming from the faint end demonstrates the most significant decline in star counts. 

In least-squares fitting, we seek to find the values of model parameters which for independent errors minimise the function $\chi^2$, given by

\begin{equation}
\chi^2=\sum_{i=1}^{N}\frac{\left( d_i - m_i \right)^2}{\sigma_i^2}
\end{equation}

\noindent where $d_i$ is the $i^{th}$ data point with an uncertainty $\sigma_i$, and $m_i$ is the corresponding point in the model. Assuming the model chosen is a good one, then the best fit should produce a $\chi^2_{min}$ which is equal to the number of degrees of freedom. If we used a single model shape, then we would require only one parameter, $\tau$, which would govern the x-axis offset between data and model. However, we do not want to specify the slope to be fitted, and so instead we use a family of templates, $\phi_s$, where $s$ is a parameter which governs the gradient of the template. In all, the template is defined over a $1^m$ range, chosen empirically as the range that is generally sufficient to show the feature we are attempting to fit, while still allowing our model to be a reasonable approximation to the feature. Since this extends over a shorter x-axis range than the complete LPD, then the y-axis range of the data that we fit to changes with $\tau$. We thus need to introduce two scaling parameters to deal with this; a vertical scale factor $k_{\tau,s}$ and a 'dc offset' $a_{\tau,s}$. For any values of $s$ and $\tau$, the relationship between the actual model fitted, $m_{\tau,s}$, and the template, $\phi_s$, is then

\begin{equation}
m_{\tau,s} = k_{\tau,s} \phi_s +a_{\tau,s}
\end{equation}

\noindent and so we are left to minimise the expression

\begin{equation}
\chi^2\left(s,\tau,k,a\right)=\sum_{i=1}^{N}\frac{\left( d_{i+\tau} - \left( k_{\tau,s} \phi_{s,i} + a_{\tau,s} \right)\right)^2}{\sigma_{i+\tau}^2}
\end{equation}

\noindent where $N$ now signifies the number of points defining our template $\phi_s$, and we have defined $\tau$ such that it corresponds to an integer number of points in the data. Simple expressions for $a_{\tau,s}$ and $k_{\tau,s}$ may be derived by looking for the values which produce a minimum in $\chi^2$. By defining our templates such that $\sum_{i=1}^{N}\phi_{s,i} = 0$ and assuming a constant error term we then find that

\begin{equation}
a_{\tau,s} = \frac{1}{N}\sum_{i=1}^{N} d_{i+\tau}
\end{equation}

\begin{equation}
k_{\tau,s} = \frac{\sum_{i=1}^{N} d_{i+\tau} \phi_{s,i}}{\sum_{i=1}^{N} \phi_{s,i}^2}
\end{equation}

\noindent A constant error term is a reasonable approximation to the error distribution of the LPD, although its actual error distribution is complicated by the covariance of neighbouring regions. Every point on the LPD is a sum of contributions from every star and so a simple Poisson distribution is not valid. However, given that there are a large number of stars contributing to every point, then the central limit theorem implies that the errors can be well approximated by a constant error in this limit. A check of the validity of this approximation and its effect on the resulting uncertainties on the parameters is conducted with the M33 data in Section 3.3, and it is shown to be robust.

Our method is then as follows: we calculate $m_{\tau,s}$ for every value of $\tau$ and $s$ by using the above prescription to calculate $k_{\tau,s}$ and  $a_{\tau,s}$. We then conduct a simple least-squares fit of these models to the LPD, in $1^m$ windows, and we find the values of $s$ and $\tau$ that minimise $\chi^2\left(s,\tau\,k_{\tau,s},a_{\tau,s}\right)$. This then gives us the TRGB. Examination of the $1\sigma$ contours in the $\tau - s$ plane also gives us the uncertainty associated with this value. Due to the general nature of this fit, it is possible that at times the method may locate a feature that is nothing to do with the TRGB, but is perhaps due to noise. However, such cases are rare for a well sampled luminosity function, and are generally obvious. They are usually easily identified by visual inspection of the fit and appropriate constraints to the parameters can be applied if neccessary. 

\subsubsection{Testing the Data Adaptive Least Squares Method}

To test the robustness of the data-adaptive slope method, we generated model CMDs that had a TRGB at 20$^{\rm{th}}$ magnitude, marked by a steady rise in star counts. Photometric noise was added by selecting a value from a gaussian distribution with $\sigma=0^m.02$, the typical photometric uncertainty in our data at 20$^{\rm{th}}$ magnitude. It should be noted that we experimented with different error weightings but found it had a negligible effect on our results, as in the range of interest around the TRGB the photometric error is well approximated by an error of $0^m.02$. We used a background of 8000 stars randomly distributed between 19$^{\rm{th}}$ and 23$^{\rm{rd}}$ magnitude and populated a red giant branch between 20$^{\rm{th}}$ and 23$^{\rm{rd}}$ magnitude. Our method was then applied to this model CMD. Measuring the S/N as the ratio of the number of stars on the TRGB to those in the background (ie. unwanted contaminants), the results as the S/N increases can be seen as the solid line in Figure 1. For a S/N greater than approximately one, the measured position of the tip differs from the actual position by, at most, $0^m.02$. This nicely shows that the method is able to identify and fit the region of the TRGB to a high accuracy.

\begin{figure}
\begin{center}
\includegraphics[width=6cm, angle=270]{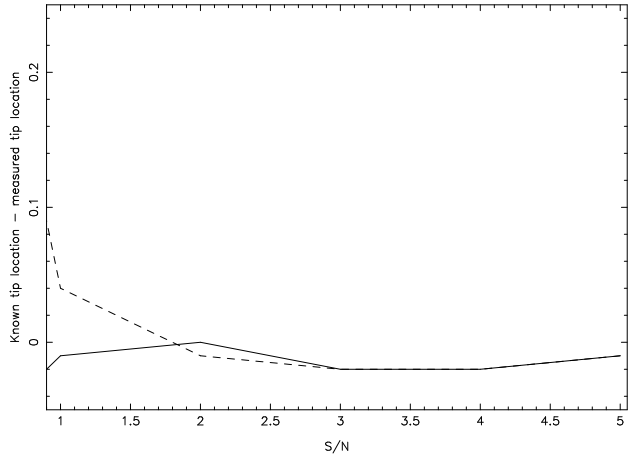}
\caption{The convergence of the measured value of the tip with its known location as the S/N is increased (solid line). Here we define the S/N to be the ratio of the number of stars on the RGB to the background contaminants. The method is accurate even at low ($\sim 1$) S/N. A second set of tests which also considers the effect of an AGB population on the algorithm shows comparable behaviour (dashed line), where this time the S/N is defined as the ratio of stars on the RGB to the sum of the background contaminants and the AGB population.}
\end{center}
\end{figure}

\begin{figure}
\begin{center}
\includegraphics[width=6cm,angle=270]{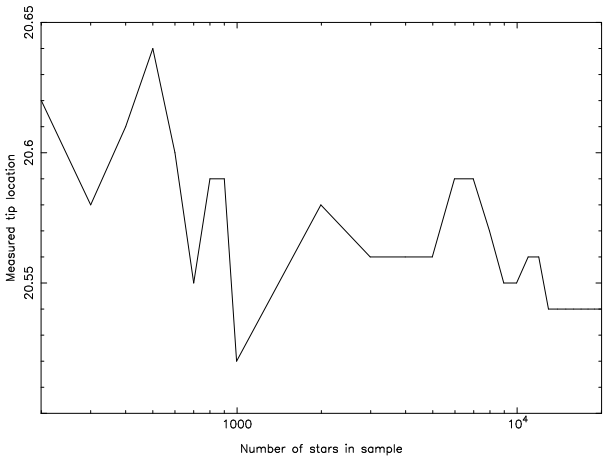}
\caption{The variation of the measured location of the TRGB with star counts in M33. Despite the stellar sample size varying by 2 orders of magnitude, the peak-to-peak variation in the measured TRGB location is only $0^m.12$ over the entire range.}
\end{center}
\end{figure}

We ran a second set of tests which also included a large population of 1000 AGB stars randomly distributed between $19^m.5$ and $20^m.5$, which will act to mask the tip (dashed line in Figure 1). The behaviour of convergence with increasing S/N remains virtually unaltered for $\rm{S/N} > 1$ even with a significant AGB population.

A third set of tests were conducted on the M33 data. In Section 3.2 we calculate that the TRGB for this galaxy lies at $20^m.54$ (uncorrected for reddening), based upon some 20000 stars contributing to the LPD. Figure 2 shows how our measured value of the TRGB varies as we continually reduce this number. It is evident that although the number of stars we use changes by 2 orders of magnitude, the peak-to-peak variation in the measured value of the tip is only $0^m.12$, decreasing to $0^m.05$ when more than 2000 stars are considered. This demonstrates the stability of our technique over a very large range in star counts. There is also evidence of a trend such that, as we go to larger numbers, the measured location of the tip tends to brighter magnitudes. This is not to be unexpected since the larger sample of stars will ensure that the brightest RGB stars lie closer to the true tip location. This fact, and the decreasing scatter in the measured values as we go to larger numbers, shows that large number statistics are desirable for such a study. 

\subsection{Relative Increases (Heuristic)}

Finally, we have developed a second, much simpler, method to estimate the location of the TRGB. This involves monitoring the relative increase in star counts between neighbouring bins in the foreground-corrected RGB luminosity function once the counts have exceeded a threshold level. Since we are only comparing neighbouring bins, then noise effects will affect us to a significant degree. To combat this, we (non-recursively) average each bin with its two immediate neighbours. The output should show a peak near the location of the TRGB. It is very useful to have such a simple method to provide an independent sanity-check of the results from the data-adaptive least-squares technique.

\section{Distance Determinations to Some Local Group Members}

\subsection{Data}

The Wide Field Camera on the 2.5m Isaac Newton Telescope consists of a four-chip EEV 4K x 2K CCD mosaic camera that images $\simeq$ 0.29 deg$^2$ per exposure (Walton et al. 2001). Over the past 3 years we have been using this to conduct a survey of the halo and outer disk of our nearest neighbour M31. This survey now extends out to over 55 kpc in projection from the centre of this galaxy. In 2002, we extended this project to include M33 which has now been mapped out to a radius of $\sim 20$ kpc. Even the early results of these surveys were startling, with the halo of M31 showing significant and unexpected substructure; the most spectacular of these being a massive tidal stream stretching out as far as we have surveyed (Ibata et al. 2001, Ferguson et al. 2002, McConnachie et al. 2003). As part of the M31 and M33 surveys we also took fields covering two of M31's dwarf spheroidal satellites, And I \& II.

Images were taken in the equivalent of Johnson V ($V'$) and Gunn i ($i'$) bands. For both surveys, most of the data was taken in photometric conditions. The few non-photometric nights were calibrated using overlapping fields and the whole system was placed on a consistent scale using all of the field overlaps. The data for And I \& II were deliberately acquired under photometric conditions. Exposure times of 800 - 1000 s per passband allowed us to reach $i' = 23^m.5$ and $V' = 24^m.5$ (S/N $\simeq$ 5), and is sufficient to detect individual RGB stars to $M_{V'}\simeq 0^m$ and main-sequence stars to $M_{V'} \simeq -1^m$ at the distance of M31. Several fields taken in poorer conditions were reobserved and co-added as necessary to give an approximately uniform overall survey depth. For this paper, we have converted our data to the Johnson-Cousins system. The transformations employed are  ${\rm I} =i' - 0.101 \times  {\rm (V-I)}$  and ${\rm  V} = V'  + 0.005  \times {\rm (V-I)}$ and have been derived by comparison with observations of several Landolt standard fields \footnote{http://www.ast.cam.ac.uk/$\sim$wfcsur/colours.php}.

All the on-target data plus calibration frames were processed using the standard INT Wide Field Survey (WFS) pipeline supplied by the Cambridge Astronomical Survey Unit (Irwin \& Lewis 2001), which provides all the usual facilities for instrumental signature removal and astrometric and photometric calibration. Internal cross-calibration of the four CCDs is achieved at a level better than 1\% for each pointing. The overall derived photometric zero points for the entire survey are good to the level of $\pm 2\%$ in both bands. Object classification uses the observed morphological structure on the CCDs but is vulnerable at faint magnitudes to distant compact galaxies, which may be misclassified as stars. However, this does not seriously affect us here as we are generally dealing only with stars near the TRGB. For the purposes of this analysis, we concern ourselves only with those objects classified as stellar in both the $V'$ and $i'$ passbands. 

As already discussed, the Galactic foreground stellar population should be subtracted from the data to minimise contaminants. In order to compensate for this population, we use a number of outer halo fields from our M31 survey for both M33 (where there is no suitable local reference field we can use) and And I (which is sufficiently close to M31 for the foregrounds to similar). Although there is a low level M31 halo population in these reference fields, the dominant population is galactic foreground stars, thus making them suitable as reference fields and allowing a statistical correction to be applied. The computed scaling using the brighter part of the CMDs compensates not only for the difference in area but also to first order the Galactic population gradient. For And II, we are able to apply a local foreground correction (see Section 3.5). In this study we typically find that the foreground contamination in the region of the TRGB is $\sim 15$ stars/sq.deg/mag for the selection criteria that we apply.

\subsection{The Adopted Absolute I Magnitude of the TRGB}

In order to calculate distances to these galaxies, we must adopt a value for the I-band absolute magnitude of the TRGB. We use the most recent observational determination of this position, given by Bellazzini et al. (2001). These authors studied the galactic globular cluster $\omega$ Centauri using a very large photometic database and an independently derived distance estimate to this system which used a detached eclipsing binary star. They derive a value of $-4^m.04 \pm 0^m.12$ for the I-band magnitude of the TRGB at a metallicity of ${\rm [Fe/H]} \simeq -1.7$ dex, which agrees extremely well with theoretically-derived estimates of this point. Although there is a small variation of this magnitude with metallicity, it is significantly less than the quoted uncertainty for the range of metallicities that we will be considering, given the stellar systems we are analysing. As such, we adopt the value as constant. It is important to note that the uncertainty on this value as it stands would be the single biggest contributor to the uncertainties in our final derived distances, and is approximately 4 times bigger than our systematic errors. This degree of uncertainty is somewhat conservative, and the error quoted is primarily due to the uncertainty in the true distance modulus of $\omega$ Cen, which is of order $0^m.1$. We therefore adopt $M_{TRGB} = -4^m.04 \pm 0^m.05$, and note that although it is possible that the scale may be shifted by up to $0^m.1$, this will affect all of our measurements in the same way. The relative distances, to say M31, will essentially remain unchanged.

\subsection{M33}

\begin{figure}
\begin{center}
\includegraphics[width=8cm]{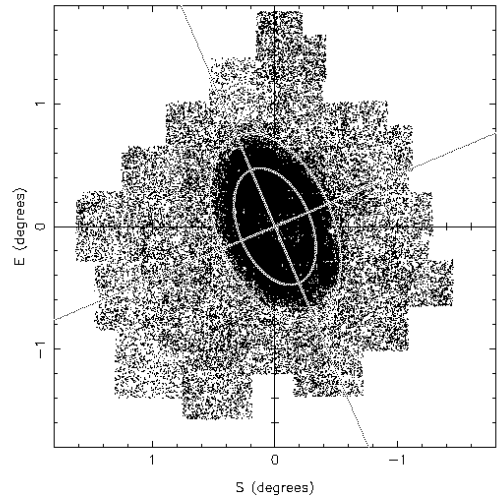}
\caption{The spatial distribution of RGB stars in M33, from our INT WFC survey of this galaxy (Ferguson et al. 2004, {\it in preparation}). Only stars brighter than ${\rm I} = 22^m.5$ lying in the strip indicated in the CMD are plotted. Stars around the edge of the disk lying in the elliptical annulus will be used in our analysis (0.72 sq. degrees). The position angle of the disk to the vertical is $\sim 23^o$.}
\end{center}
\end{figure}

\begin{figure}
\includegraphics[width=8cm,]{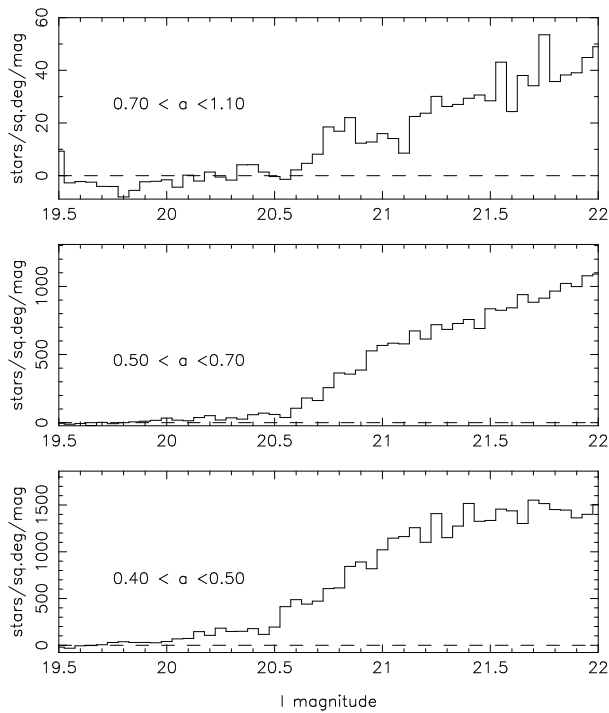}
\caption{RGB luminosity functions for M33 as a function of galactocentric radius. Each RGB luminosity function is created from stars within elliptical annulii with semi-major axes of $a^\circ$ and have been foreground-corrected. The innermost RGB luminosity function may suffer from stellar blending. However, the similarity of the outer two RGB luminosity functions suggest that blending is having a negligible effect on our TRGB measurement for M33, which uses stars located at $0^\circ.5 < a < 0^\circ.8$.}
\end{figure}

\begin{figure}
\begin{center}
\includegraphics[width=8cm]{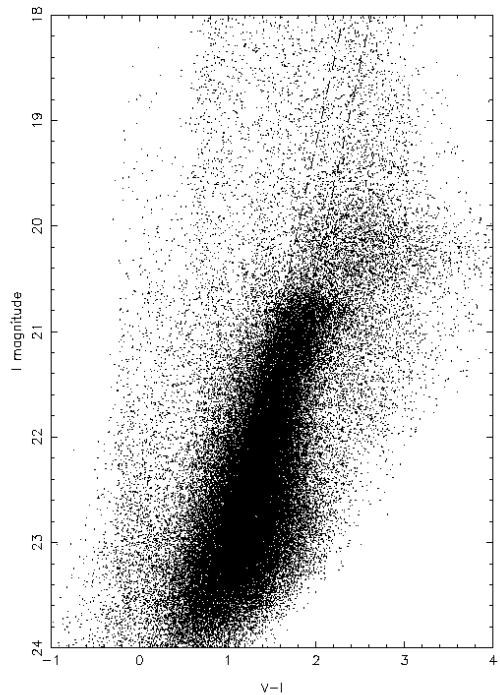}
\caption{The colour magnitude diagram of the spiral galaxy M33, using stars in the elliptical annulus shown in Figure 3. A bright AGB star population is obvious above the TRGB. Only stars in the strip indicated were used in the construction of the luminosity function. Selecting such a narrow strip along the RGB still gives us ample stars for our analysis and significantly decreases the AGB contamination.} 
\label{m33}
\end{center}
\end{figure}

\begin{figure}
\begin{center}
\includegraphics[width=8cm]{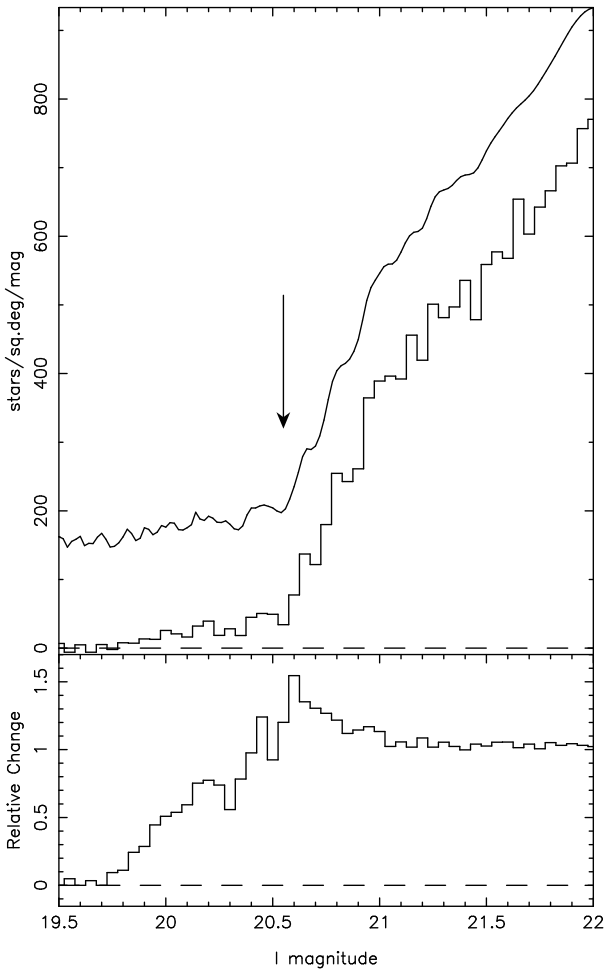}
\caption{Upper panel: The RGB luminosity function for M33 (lower curve is binned luminosity function, upper curve is offset LPD) with an arrow indicating the measured position of the TRGB using the fit of an adaptive slope. This location is partially masked by an AGB population, although its determination is still possible by use of the least squares method. The area of sky used in this analysis is the area contained by the elliptical annulus shown in Figure 3. Lower panel: The TRGB as identified by the heuristic method.}
\end{center}
\end{figure}

The RGB distribution from the central pointings of our survey of M33 (Ferguson et al. 2004, {\it in preparation}), located at 1h 33m 50.9s, 30$^\circ$ 39' 35.8'', is shown in Figure 3 for all stars brighter than ${\rm I} = 22^m.5$ in the strip indicated in Figure 5. The orientations of M33's axes are indicated, as is an elliptical annulus around the edge of the disk, selected so as to avoid the effects of crowding in the inner region while still sampling stars across the entirety of M33's disk. The inner ellipse has a semi-major axis $a = 0^\circ.5$, while the outer ellipse has $a = 0^\circ.8$. To provide a check that our resulting RGB luminosity function does not suffer heavily from blending effects, we have compared the RGB luminosity functions for elliptical annuli at different galactocentric radius. The results are shown in Figure 4 and are all foreground-corrected. The outer annulus samples a region of M33 where we expect negligible crowding, while the innermost annulus might be expected to sample many blended stars. Indeed, the innermost RGB luminosity function is seen to extend to brighter magnitudes than the outermost RGB luminosity function suggesting that blending may be affecting the inner fields to some extent. In this respect, the middle annulus results in a RGB luminosity function whose overall shape is very similar to the outermost RGB luminosity function, given that the outer RGB luminosity function samples fewer stars and is necessarily more noisy. In fact, the location of the TRGB for the outer annulus is at $20^m.56 \pm 0^m.02$, which compares very well with our final derived value of $20^m.54 \pm 0^m.01$. This implies that the effect of crowding on our M33 data is negligible for $a > 0^\circ.5$. 

The CMD of the stars we analyse to measure the TRGB are shown in Figure 5. It is obvious from the width of the RGB that M33 contains a larger spread in age/metallicity than the Andromeda dwarfs (Figure 8), and the region of the tip lacks a definite edge with many bright AGB stars visible in that region. This is illustrated in the RGB luminosity function (Figure 6) where we see that the tip of the luminosity function delimits a change in slope rather than an edge, and the AGB stars are also visible here for ${\rm I} < 20^m.6$. A weak blue plume representative of a younger stellar population is also evident in the CMD at ${\rm V - I} \simeq 0$.

A least-squares fit to the RGB LPD gives a TRGB of $ {\rm I} = 20^m.54 \pm 0^m.01$. The uncertainty in this measurement is given by the $1\sigma$ contour in the $\tau - s$ plane (see Figure 7). This contour plot was constructed assuming a constant error term in the LPD (Section 2.2). To provide a check of this approximation, we also conduct a similar $\chi^2$ fit to the binned luminosity fuction, where we know the errors to be well described by a Poisson distribution. The expressions for $a$ and $k$ given by Equations 3 and 4 become necessarily more complex, but the contour levels remain nearly unchanged. 

From the extinction maps of Schlegel et al. (1998), the interstellar reddening $\rm{E(B - V)}$ is found to be $0^m.042$, and they state that these values are accurate to within 16\%. Using the relation between reddening and extinction in the Landolt system (Landolt 1992), given in the same paper, $A_I = 1.94 E\left(B - V\right)$, implies that $A_I = 0^m.081$. We take the absolute I magnitude of the TRGB to be $M_{TRGB} = -4^m.04 \pm 0^m.05$ as discussed in Section 3.2.. The other uncertainty that we take into account are the mean systematic photometric error in our measurements, which is of order $0^m.02$. Blending has already been shown to have a negligible effect on our measurement of the TRGB. Combining the observational errors lead to an overall error of $0^m.03$ - $0^m.04$ in the measured value of the TRGB magnitude. We calculate the distance, $D$, to the galaxy from $I - M_{TRGB} = 5\log_{10}D - 5 + A_I$ and conclude that overall, $D_{M33} = 794 \pm 23$ kpc. Application of the heuristic method identifies a tip at $20^m.6$, in good agreement with the more detailed treatment.

This result brings the TRGB distance to M33 into excellent agreement with the Cepheid distances (see Table 1). The suggestion of a significant amount of reddening in M33 by Kim et al (2002), to explain the discrepancy between their TRGB measurements and their Cepheid measurements (Lee et al. 2002), seems unlikely. We attribute the difference in our results to the fact that we have better statistics than was possible to achieve with their smaller HST fields - we do not see evidence for a significant jump at $\sim 20^m.9$, as these authors do (see their Figure 4). Of the 10 fields that they analyse, several of the luminosity functions do show evidence for a discontinuity in the slope at $\sim 20^m.5 - 20^m.6$, which may well be associated with the discontinuity that we attribute as the TRGB in our luminosity function. Since our algorithm defines the location of the TRGB somewhat differently to an edge detection algorithm, then this location is highlighted in preference to any absolute change in counts. Additionally, it is unlikely that we have instead measured a feature due to AGB contamination - an AGB population is clearly observed in the CMD and RGB luminosity function for this galaxy (Figures 5 and 6) and so there would have to be a second AGB population in the CMD exactly coincident with the top of the RGB, although this seems unlikely. We would also expect that, if this was the case, their effect would be lessened in a luminosity function located at larger galactocentric distance. Inspection of Figure 4 shows that this is not the case for these stars, even though the contribution from the known AGB component can be seen to reduce in the expected way.

Kim et al. has also calculate the distance to M33 using the red clump (RC), finding agreement with their TRGB measurement. This feature represents stars with a helium core and a hydrogen shell passing through a relatively slow phase of their luminosity evolution, leading to a clump appearing in the CMD. They have recently been suggested for use as a distance indicator by Paczynski \& Stanek (1998). They compared the I-band magnitude of OGLE RC stars seen through Baade's Window with a sample from the Hipparcos catalogue with parallaxes known to $\le 10\%$, allowing the absolute I magnitude of the RC ($M^{RC}_I$), and hence a distance, to be calculated for the stars seen towards the Galactic Centre. The reliability of this method when applied to other stellar systems relies on the similarity of the Hipparcos sample with the RC stars in the system of interest, and remains a controversial point due to possible dependecies of $M^{RC}_I$ on the properties of the population. Theoretical models seem to predict a relatively strong dependency of $M^{RC}_I$ with age and metallicity (eg. Cole 1998; Girardi et al. 1998; Girardi 2000; Girardi \& Salaris 2001) while observational studies are in apparent disagreement with each other (eg. Udalski 1998; Sarajedini 1999). However, both observation and theory show that $M^{RC}_I$ is significantly fainter for stars with ages $> 10$ Gyrs than for their intermediate age counterparts (Udalski 1998, Girardi \& Salaris 2001). Kim et al (2002) assume that the majority of the stellar population in M33 is old ($> 10$ Gyrs) in order to associate the mean metallicity of the RGB with the metallicity of the RC. This metallicity is then used to calculate $M^{RC}_I$ using the calibrations for $M^{RC}_I$ vs $[Fe/H]$ by Udalski (1998) and Popowski (2000). However, it is not clear that these calibrations are valid for M33 given the assumption about the age of the bulk of the stellar population. If $M^{RC}_I$ is significantly fainter in this galaxy due to age then M33 will be correspondingly closer than these authors calculated.

\begin{figure*}
\includegraphics[width=8cm,angle=270]{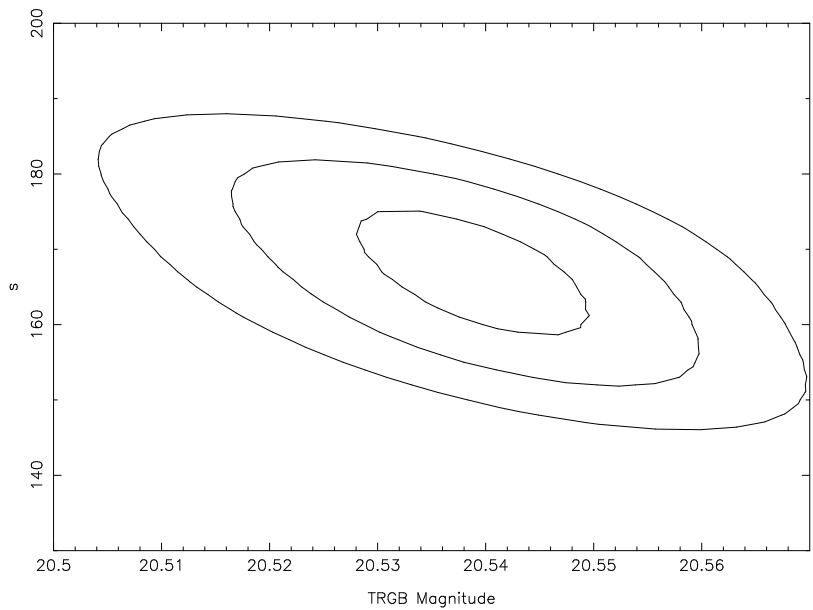}
\caption{Contour plot of the measured TRGB location versus the slope parameter, $s$, for M33. Contours show the 1, 2 and 3$\sigma$ confidence levels.}
\end{figure*}

\subsection{And I}

\begin{figure*}
\includegraphics[width=12cm,angle=270]{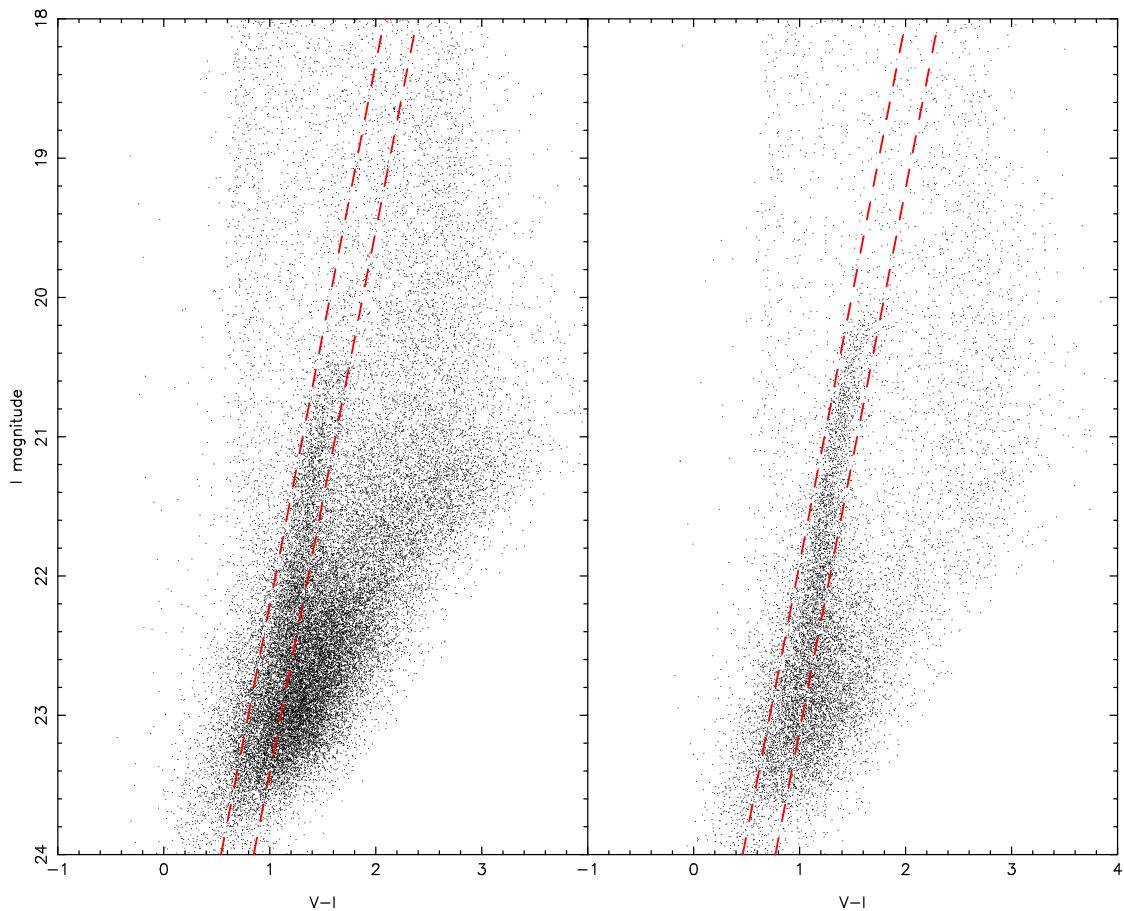}
\caption{The colour magnitude diagrams for the dwarf galaxies And I (left panel) and And II (right panel). The luminosity function is constructed from the region contained within the dotted lines. The RGB is easily visible in both cases with the tip at $I \simeq 20^m.45$ for And I and $I \simeq 20^m.1$ for And II. The additional red component centred around $\rm{V-I} \simeq 2.4$ in the left panel is due to the giant Andromeda tidal stream discovered by our INT survey of M31 (Ibata et al 2001, Ferguson et al. 2002, McConnachie et al. 2003) which extends over the region around And I.}
\end{figure*}

\begin{figure}
\begin{center}
\includegraphics[width=8cm]{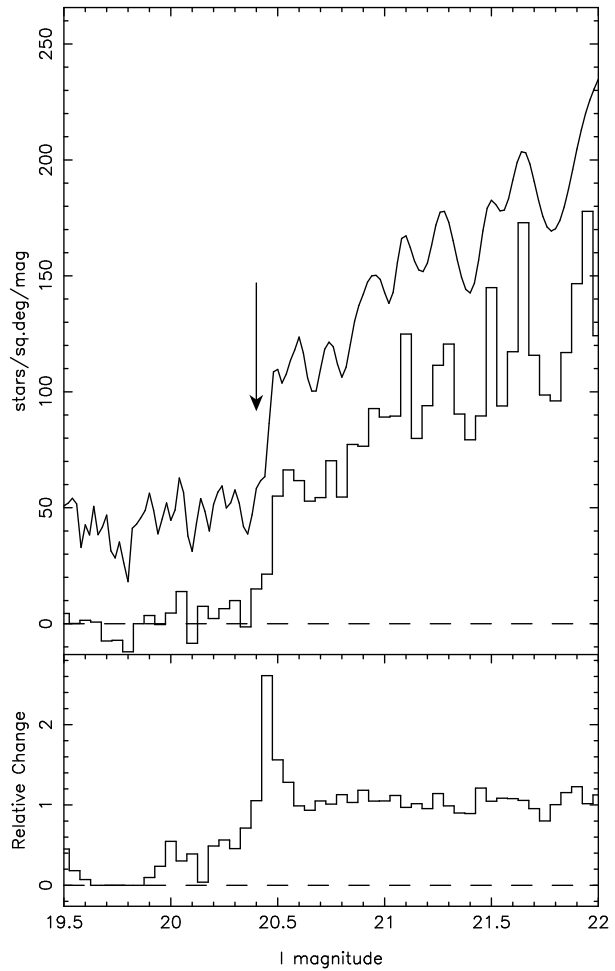}
\caption{Upper panel: The solid line shows the (RGB) luminosity function of And I. The dashed line shows the (RGB) LPD. The total are of sky surveyed was 0.47 sq. degrees. The arrow indicates the position that we measure the TRGB to be located using our adaptive slope. Lower panel: Measurement of the location of the tip using the heuristic method, showing excellent agreement with the adaptive slope.}
\end{center}
\end{figure}

\begin{figure}
\begin{center}
\includegraphics[width=8cm]{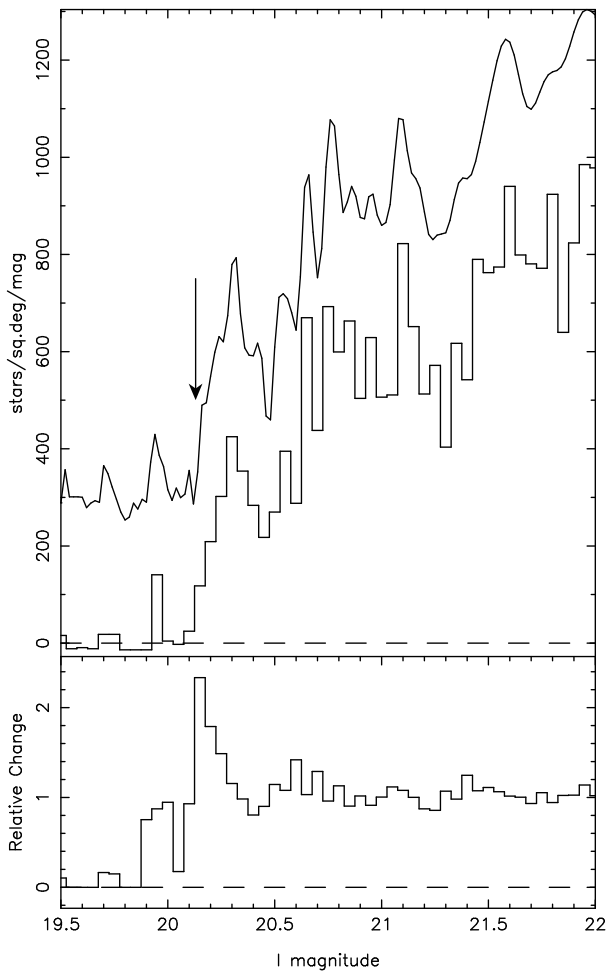}
\caption{Same as Figure 9, but for the dwarf galaxy And II. This shows all stars located within $0.1^\circ$ of the centre of this galaxy, allowing the rest of the field to be used as a local foreground correction.}
\end{center}
\end{figure}

The dwarf spheroidal galaxy And I was discovered by van den Bergh (1971) and lies at 0h 45m 39.8s, 38$^\circ$ 2' 28''. It is a satellite galaxy of M31 at a projected distance of $\sim 40$ kpc. Mould \& Kristian (1990) were the first to study its RGB using CCD images taken at the prime focus of the Hale telescope. Using an eyeball determination of the TRGB they found it to be at a distance of $790 \pm 60$ kpc. They also concluded the galaxy to be of intermediate metallicity. Da Costa et al. (1996) obtained HST WFPC2 images of this system and by comparison with the giant branches of several globular clusters confirmed the metallicity to be \mbox{$<[Fe/H]> = -1.45 \pm 0.2$ dex}. By analysing the morphology of the horizontal branch they also came to the conclusion that the majority of the population is approximately 10 Gyrs old, However, the presence of a blue horizontal branch and RR Lyrae stars shows that there is a minority population at least 3 Gyrs older than this. And I, like many other dwarf spheroidals, appears to have undergone an extended period of star formation.

The colour magnitude diagram for And I is shown in the left panel of Figure 8. The upper few magnitudes of the RGB are clearly observed and even by visual inspection we can determine that the tip is at $I \simeq 20^m.45$. One interesting point of note is the additional red feature ($\rm{V - I} > 2$) that is not present in the And II CMD (right panel). This is due to the giant stellar stream first discovered by our survey of the outer regions of M31, which is sufficiently broad that part of it lies along the line-of-sight to And I.

Application of the least-squares method identifies the TRGB to be at $I = 20^m.40^{+0^m.03}_{-0^m.02}$, as shown in Figure 9. Schlegel et al. (1998) gives $E(B - V) = 0^m.056$ for the interstellar reddening, and so we derive $D_{AndI} = 735 \pm 23 $ kpc, closer than had previously been determined (Table 2) but still in reasonable agreement. Application of the heuristic method identifies the tip to lie at $20^m.45$ (Figure 9), in excellent agreement with the adaptive slope, not unexpected for such a clean LPD.

\begin{table*}
\begin{minipage}{115mm}
\begin{tabular*}{110mm} {@{\extracolsep{\fill}}c c c c}
\label{tabm33}
\\${\rm (m - M)_o}$ & Distance & Method & Reference \\
\hline\\
$24.50 \pm 0.06$ & $794 \pm 23$ kpc & TRGB & This study \\
$24.52 \pm 0.14$ & $802 \pm 51$ kpc & Cepheids & Lee et al. 2002 \\
$24.81 \pm 0.13$ & $916 \pm 55$ kpc & TRGB & Kim et al. 2002 \\ 
$24.80 \pm 0.14$ & $912 \pm 59$ kpc & Red Clump & Kim et al. 2002\\
\hline\\
\end{tabular*}
\caption{The distance to M33, along with a selection of the most recent distance estimates to M33 based on HST data.}

\begin{tabular*}{110mm}{@{\extracolsep{\fill}}c c c c}
\\${\rm (m - M)_o}$ &Distance & Method & Reference  \\
\hline\\
$24.33 \pm 0.07$ & $735 \pm 23$ kpc & TRGB & This study \\
$24.50 \pm 0.15$ & $790 \pm 60$ kpc & TRGB & Mould \& Kristian 1990 \\
$24.55 \pm 0.08$ & $810 \pm 30$ kpc & Horizontal Branch stars & Da Costa et al. 1996 \\
\hline\\
\end{tabular*}
\caption{The distance to And I, together with previous estimates.}

\begin{tabular*}{110mm}{@{\extracolsep{\fill}}c c c c}
\\${\rm (m - M)_o}$ & Distance & Method & Reference  \\
\hline\\
$24.05 \pm 0.06$ & $645 \pm 19$ kpc & TRGB & This study \\
$23.83^{+0.46}_{-0.38}$ & $583^{+124}_{-103}$ kpc & Giant branch fitting & Koenig et al. 1993 \\
$24.17 \pm 0.06$ & $680 \pm 20$ kpc & Horizontal Branch stars & Da Costa et al. 2000 \\
\hline\\
\end{tabular*}
\caption{The distance to And II, together with previous estimates.}
\end{minipage}
\end{table*}

\subsection{And II}

And II is another dwarf spheroidal satellite of M31 discovered at the same time as And I. It lies at 1h 16m 29.8s, 33$^\circ$ 25' 9'', approximately 140 kpc in projection from M31, towards M33. Despite the fact that it lies significantly closer to M33 in projection, the larger mass of M31 means that it is most likely a satellite of M31. Da Costa et al. (2000) derive a mean metallicity of $<[Fe/H]> = -1.49 \pm 0.11$ dex with a significantly larger abundance spread than in And I. An old ($\geq 10$ Gyrs) population is again implied from the presence of a blue horizontal branch. The presence of young or intermediate populations in this galaxy remains a possibility from their study, but we detect no sign of such a population to $M_V \simeq 0^m$.

The colour magnitude diagram for And II (right panel of Figure 8) again shows the upper parts of the RGB very clearly, with a tip visible at $I \simeq 20^m.1$. We construct the luminosity function for And II by using stars withing $0.1^\circ$ from its centre. This then allows us to apply a local foreground correction to the luminosity function by using as a reference population all stars on the same INT field outside of this radius, scaled in the usual way. The result is shown in Figure 10, where an unambiguos tip can be observed. We find the TRGB to lie at $20^m.13 \pm 0^m.02$ ($20^m.15$ using the heuristic technique). From Schlegel et al. (1998), $E(B-V) = 0^m.063$. This corresponds to $D_{AndII} = 645 \pm 19$ kpc, in reasonable agreement with estimates that have been made using other methods (Table 3). 

\section{Summary}

In this paper we have introduced a new quantitative method for the determination of distances to metal poor stellar systems using the TRGB as a standard candle, accurate to typically $\pm 0^m.02$ rms, $\pm 0^m.03$ systematic. The method involves a least-squares fit of a data-adaptive slope template to the foreground-corrected RGB LPD in $1^m$ windows, to find the region of the LPD that is best fit as a simple slope function ie. the region that shows the most significant change in slope as we go to brighter magnitudes.  This determines the apparent magnitude of the TRGB, as represented in the luminosity function. Observations and stellar evolutionary codes agree that this magnitude is virtually constant for old, metal poor populations and so the photometric parallax can easily be calculated.

We believe that our method provides significant improvements over previous attempts to locate the TRGB in a luminosity function. By only considering stars along a strip in the CMD, we minimise contamination from other stellar populations. Further, for systems with broader red giant branches we ensure our sample of RGB stars all have similar age/metallicity rather than analysing all the stars in the RGB. Poisson noise will affect us to a much smaller degree than previous methods since we consider a range in magnitude when determining the position of the tip, rather than just comparing neighbouring star counts. Tests of our method on both synthetic and real data demonstrate the robustness of this method, even with relatively poor S/N. Perhaps most importantly, we demonstrate that accurate TRGB determinations require the use of Wide Field Cameras, such as that on the INT, in order to boost star counts and allow the measured RGB to be as accurate and complete a representation of the population as possible. 

The methodology that we develop in this paper is applied to three galaxies - And I \& II, and M33. We derive distances to these objects of $735 \pm 23$ kpc, $645 \pm 19$ kpc and $794 \pm 23$ kpc respectively (see Tables 1, 2 and 3 for comparison with previous results). The contribution to this uncertainty due to the algorithm is only of order $15$ kpc, demonstrating a significant improvement over other previous techniques that have been used. What is perhaps of greatest importance when comparing this algorithm with, for example, the edge detection alogorithm of Sakai et al. (1996) or the maximum likelihood method of Mendez et al. (2002), is that it uses a different set of criteria to define the location of the tip. In the majority of situations, where the tip position is marked by a significant jump in star counts, the methods will be expected to give the same result. In other situations (such as a gradual increase in star-counts at the TRGB edge, nicely demonstrated in M33) we would not necessarily expect the algorithms to agree as they search for fundamentaly different features in the luminosity function. In the second paper in this series (McConnachie et al. 2004) we apply our algorithm to 11 other Local Group galaxies - And III, And V, And VI, And VII, NGC185, NGC 147, Pegasus, WLM, LGS3, Cetus and Aquarius - to create a homogeneous set of distances to nearby galaxies, a majority of which are members of the M31 subgroup.

\section*{Acknowledgments}

Many thanks to the referee, Barry Madore, for a careful reading of this manuscript and useful comments. Additionally, we would like to thank Minsun Kim and collaborators for supplying us with their M33 HST/WFPC2 data, and Eline Tolstoy for her opinion on some of the arguments presented in this paper. The research of AMNF has been supported by a Marie Curie Fellowship of the European Community under contract number HPMF-CT-2002-01758.


\begin{thebibliography}{}

\bibitem[]{} Bellazzini, M., Ferraro, F.R., Pancino, E., 2001, ApJ, 556, 635

\bibitem[]{} Cole, A.A., 1998, ApJL, 500, 137

\bibitem[]{} Crocker, D.A., Rood, R.T., 1984, IAU Symp. 105, p159

\bibitem[]{} Da Costa, G. S., Armandroff, T. E., 1990, AJ, 100, 162

\bibitem[]{} Da Costa, G. S., Armandroff, T. E., Caldwell, N., Seitzer, P., 1996, AJ, 112, 6

\bibitem[]{} Da Costa, G. S., Armandroff, T. E., Caldwell, N., Seitzer, P., 2000, AJ, 119, 705

\bibitem[]{} Ferguson, A. M. N., Irwin, M., Ibata, R., Lewis, G.,Tanvir, N., 2002, AJ, 124, 1452

\bibitem[]{} Girardi, L., 2000, Highlights Astron., 12

\bibitem[]{} Girardi, L., Groenewegen, M.A.T., Weiss, A., Salaris, M., 1998, MNRAS, 301, 149

\bibitem[]{} Girardi, L., Salaris, M., 2001, MNRAS, 323, 109

\bibitem[]{} Harris, W.E., Durrell, P.R., Pierce, M.J., Secker, J., 1998, Nature, 395, 45

\bibitem[]{} Ibata, R., Irwin, M., Lewis, G., Ferguson, A. M. N., Tanvir, N., 2001, Nature, 412, 49 

\bibitem[]{} Iben, I. J., Renzini, A., 1983, ARA\&A, 21, 271

\bibitem[]{} Irwin, M., Lewis, J., 2001, NewARev., 45, 105

\bibitem[]{} Kim, M., Kim, E., Lee, M.G., Sarajedini, A., Geisler, D., 2002, AJ, 123, 244 

\bibitem[]{} Koenig, C.B.H., Nemec, J.M., Mould, J.R., Fahlman, G.G., 1993, AJ, 106, 1819

\bibitem[]{} Landolt, A.U., 1992, AJ, 104, 340L

\bibitem[]{} Lee, M. G., Freedman, W. L., Madore, B. F., 1993, ApJ, 417, 553

\bibitem[]{} Lee, M.G, Kim, M., Sarajedini, A., Geisler, D., Gieren, W., 2002, ApJ, 565, 959

\bibitem[]{} Madore, B.F., Freedman, W.L., 1995, AJ, 109, 1645

\bibitem[]{} Madore, B.F., McAlary, C.W., McLaren, R.A., Welch, D.L., Neugebauer, G., Matthews, K., 1985, ApJ, 294, 560

\bibitem[]{} Mateo, M., 1998, ARA\&A, 36, 435 

\bibitem[]{} McConnachie, A.W., Irwin, M.J., Ibata, R.A., Ferguson, A.M.N., Lewis, G.F., Tanvir, N., 2003, MNRAS, 343, 1335

\bibitem[]{} McConnachie, A.W., Irwin, M.J., Ibata, R.A., Ferguson, A.M.N., Lewis, G.F., Tanvir, N., 2004, MNRAS, {\it submitted}

\bibitem[]{} Mendez, B., Davis, M., Moustakas, J., Newman, J., Madore, B.F., Freedman, W.L., 2002, AJ, 124, 213

\bibitem[]{} Merritt, D., Ferrarese, L., Joseph, C. L.,  2001, Science, 293, 1116 

\bibitem[]{} Metcalfe, N., Shanks, T., 1991, MNRAS, 250, 438

\bibitem[]{} Mould, J., Kristian, J.,  1990, ApJ, 354, 438

\bibitem[]{} Paczynski, B., Stanek, K.Z., 1998, ApJL, 494, 219

\bibitem[]{} Popowski, P., 2000, ApJL, 528, 9

\bibitem[]{} Sakai, S., Madore, B. F., Freedman, W. L, 1996, ApJ, 461, 713

\bibitem[]{} Salaris, M., Cassisi, S., 1997, MNRAS, 289, 406

\bibitem[]{} Sarajedini, A., 1999, AJ, 118, 2321

\bibitem[]{} Schlegel, D. J., Finkbeiner, D. P., Davis, M., 1998, AJ, 500, 525

\bibitem[]{} Tanvir, N., 1999, ASP Conf.Ser. 167, Eds. D. Egret and A. Heck, p84

\bibitem[]{} Udalski, A., 1998, Acta Astron., 48, 383

\bibitem[]{} van den Bergh, S., 1971, ApJL, 171, L31

\bibitem[]{} Walton, N. A., Lennon, D. J., Greimel, R., Irwin, M., Lewis, J., Rixon, G. T., 2001, ING Newsl., No.4, 7

\bibitem[]{} Zoccali, M., Piotto, G., 2000, A\&A, 358, 943

\end{thebibliography}
\end{document}